\documentclass[prb,aps,amssymb,twocolumn]{revtex4-1}
\usepackage{amsmath}
\usepackage{amssymb}
\usepackage{amsthm}
\usepackage{amsfonts}
\usepackage{listings}
\usepackage{enumerate}
\usepackage{latexsym}
\usepackage{color}
\usepackage[colorlinks=true,urlcolor=blue,citecolor=blue,linkcolor=blue]{hyperref}
\usepackage{psfrag}

\usepackage{bm}
\usepackage{graphicx}
\usepackage{subfigure}

\newcommand{\beq}{\begin{equation}}
\newcommand{\eneq}{\end{equation}}

\newcommand{\bal}{\begin{align}}
\newcommand{\eal}{\end{align}}

\def\ie{{\it i.e.},\ }

\input{epsf}

\begin{document}

\tolerance 10000

\newcommand{\vk}{{\bf k}}

\title{Topological phases in the $\text{TaSe}_3$ compound}

\author{Simin Nie$^{1}$}
\author{Lingyi Xing$^{2}$}
\author{Rongying Jin$^{2}$}
\author{Weiwei Xie$^{3}$}
\author{Zhijun Wang$^{4}$}
\email{zjwang@princeton.edu}
\author{Fritz B. Prinz$^{1}$}

\affiliation{${^1}$Department of Materials Science and Engineering, Stanford University, Stanford, California 94305, USA}
\affiliation{${^2}$Department of Physics and Astronomy, Louisiana State University, Baton Rouge, LA, 70803, USA}
\affiliation{${^3}$Department of Chemistry, Louisiana State University, Baton Rouge, LA, 70803, USA}
\affiliation{${^4}$Department of Physics, Princeton University, Princeton, New Jersey 08544, USA}

\date{\today}

\begin{abstract}
Based on first-principles calculations, we show that stoichiometric TaSe$_3$, synthesized in space group $P2_1/m$, belongs to
a three-dimensional (3D) strong topological insulator (TI) phase with $Z_2$ invariants (1;100). The calculated surface spectrum
shows clearly a single Dirac cone on surfaces, with helical spin texture at a constant energy contour. To check the stability
of the topological phase, strain effects have been systematically investigated, showing that many topological phases survive in
a wide range of the strains along both the a- and c-axes, such as strong TI (STI), weak TI (WTI) and Dirac semimetal phases.
TaSe$_3$ provides us an ideal platform for experimental study of topological phase transitions. More interestingly, since
superconductivity in TaSe$_3$ has been reported for a long time, the co-existence of topological phases and superconducting
phase suggests that TaSe$_3$ is a realistic system to study the interplay between topological and superconducting phases in the future.
\end{abstract}
\maketitle

\section{introduction}
The layered transition-metal trichalcogenides MX$_3$ (M=Nb, Ta; X=S, Se) have attracted lots of interest because of the appearance
of the charge-density-wave (CDW) states at low temperature (LT)\cite{tsutsumi1977direct,wilson1979bands,cava1985low,srivastava1992preparation}.
For instance, both TaS$_3$\cite{cava1985low} and NbSe$_3$\cite{ekino1987electron} are metals at room temperature (RT), and undergo two
different CDW transitions as decreasing temperature. But TaSe$_3$ is an exception, in which no CDW transition has been found
yet\cite{haen1978low}. Instead, TaSe$_3$ remains semimetallic from RT to LT and becomes superconducting at T$_c=2.3$ K
\cite{sambongi1977superconductivity,yamamoto1978sc,nagata1989superconductivity}. The earlier theoretical studies\cite{canadell1990comparison}
suggest that it could be either a semiconductor or a semimetal, depending on the relative energy level of an anti-bonding Se $p$ band
to that of the Ta $d_{z^2}$ band. So far, first-principles calculations of TaSe$_3$ have not been reported yet, and the electronic
structures, Fermi surfaces and topological properties are still unrevealed.

Meanwhile, topological superconductors have attracted much interest due to the emergence of Majorana
fermions\cite{wilczek2009majorana,leijnse2012introduction} and the potential application in quantum
computation\cite{kitaev2003fault}. The previous work by Fu {\it et al.}\cite{fu2008superconducting}
has proposed that the topological superconductivity can be realized on the interface between a TI
and a BCS superconductor by proximity effect\cite{yan2013large,wang2015topological,wu2015cafeas,schoop2015dirac,xu2016topological}.
Very recently, topological superconductivity has been observed on the surface of iron-based
superconductors below T$_c$\cite{zhang2017observation}, which suggests that a more promising
approach to engineer TSCs is to propose superconducting materials with non-trivial topology
in their electronic structures\cite{wang2015topological}. Herein, this approach would avoid
the structural compatibility and overcome the fabrication challenges related to the interfaces
or heterostructures\cite{wu2015cafeas,wang2015topological,schoop2015dirac,xu2015two,xing2016superconductivity,xu2016topological,wang2016spontaneous}.
However, most superconductors do not have non-trivial bulk topology in electronic structures\cite{das2012zero,yan2013large},
and some topological candidates do need doping to induce superconductivity, like Cu$_x$Bi$_2$Se$_3$\cite{hor2010superconductivity,wray2010observation},
FeTe$_x$Se$_{1-x}$\cite{wang2015topological,zhang2017observation}, {\em etc.} Therefore, it is challenging
to find a superconductor with non-trivial electronic topology.

In this work, based on first-principles calculations, we show that the single crystal
$\text{TaSe}_3$, known as a superconductor for many
years\cite{sambongi1977superconductivity,yamamoto1978sc,nagata1989superconductivity}, has
non-trivial electronic structure. The crucial band inversion happens at B point even
without spin-orbit coupling (SOC).
This is different from the situation in Bi$_2$Se$_3$\cite{zhang2009topological,zhang2010first},
in which the band inversion is due to the strong SOC effect.
Our detailed analysis indicates that the band inversion is attributed to the ``broken" type II chains, 
especially the Se3-Ta1 and Se6-Ta2 bonds.
To shorten these bonds by compressive strains would enlarge the band gap and remove the band inversion.
Inclusion of SOC in TaSe$_3$ opens a continuous direct gap in the entire Brillouin zone (BZ),  but doesn't change the energy ordering of the bands at the time-reversal invariant momentum (TRIM) points.
The $Z_2$ invariants ($\nu_0$;$\nu_1$,$\nu_2$,$\nu_3$)\cite{hasan2010colloquium,qi2011topological}
are calculated to be (1;100). The strong topological index $\nu_0=1$ guarantees the existence of the Dirac-cone states on surfaces\cite{fu2007topological}, which has
been further confirmed by our surface calculations. To check the stability of the
topological phase, strain effects have been systematically investigated, showing
that many topological phases survive in a wide range of the strains along both
the $a$- and $c$-axes, such as STI, WTI and Dirac semimetal phases. It's an ideal
platform for experimental study of topological phase transitions. As it becomes
superconducting below T$_c$, TaSe$_3$ also provides us a realistic system to
investigate the interplay between the topological surface states and superconductivity.

This paper is organized as follows. In Sec. II we will
introduce the details of first-principles calculations. In Sec. III,
the calculation results are presented. Finally, Sec. IV contains the discussion and conclusion.
\begin{figure}[tp]
\includegraphics[width=3.5in]{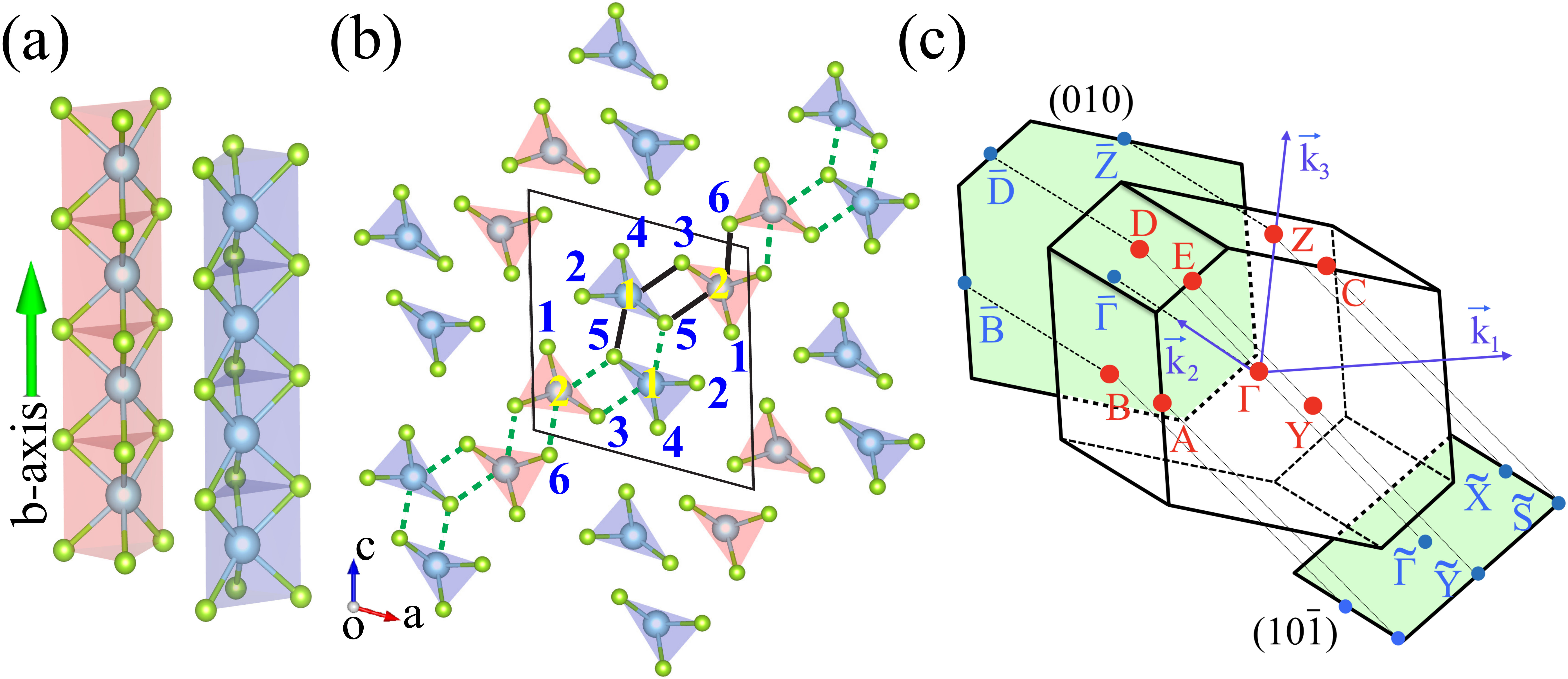}
      \caption{(Color online) Crystal structure and BZs of TaSe$_{3}$. (a) The side view of the type I and type II chains colored in blue and red, respectively.
      Two neighbor prisms are differed by $b/2$ in the $b$-direction, making the Ta atoms in the planes of the Se$_3$ triangles of the neighbor prism.
(b) The projection view along the chain direction (\ie the $b$-direction). The primitive unit cell is shown in light black parallelogram. The nonequivalent Ta atoms (large balls) and Se atoms (small balls) in the primitive unit cell are labeled by yellow and light blue numbers, respectively. The bonds (green dashed lines) of Se5-Ta1-Se3 and Se5-Ta2-Se6 make the prisms form a layer (spanned by the $b$-axis and $(a+c)$-axis).  (c) Bulk BZ, projected surface BZs, and high-symmetry points.}
\label{structure}
\end{figure}

\section{calculation method}

The first-principles calculations were performed within the framework of full-potential linearized-augmented
plane-wave (FP-LAPW) method implemented in WIEN2K simulation package\cite{blaha2002wien2k}. Modified Becke-Johnson
exchange potential together with local density approximation for the correlation potential was used to obtain
accurate band structures\cite{tran2009accurate}. SOC was included as a second variational step self-consistently.
The radii of the muffin-tin sphere (RMT) were 2.5 Bohr for Ta and 2.38 Bohr for Se, respectively. The
$k$-points sampling grid of the BZ in the self-consistent process was 7 $\times$ 19 $\times$ 6.
The truncation of the modulus of the reciprocal lattice vector $K_{max}$, which was used for the expansion of
the wave functions in the interstitial region, was set to $R_{MT} \times K_{max} = 7$. The geometry optimization
including SOC interaction was carried out within the framework of the projector augmented wave (PAW) pseudopotential
method implemented in Vienna $ab~ initio$ simulation package (VASP)\cite{kresse1996_1,kresse1996_2}.
The ionic positions were relaxed until force on each ion was less than 0.005 eV ${\AA}^{-1}$. PHONOPY was employed to
calculate the phonon dispersion through the DFPT method \cite{togo2008first}.

\section{results}
$\text{TaSe}_3$ crystallizes in the monoclinic layered structure with space group $P2_1/m$. The basic building blocks of $\text{TaSe}_3$ are
parallel trigonal-prismatic chains along the $b$-axis, as shown in Fig.~\ref{structure}(a). Each chain is made by a linear stacking of irregular prismatic cages, which consists of
six selenium atoms at the corners and one tantalum atom at the center. In the top view of Fig.~\ref{structure}(b), these chains are classified as
type I (\ie $d_{s}^I=d_{24}$=2.57\AA) and type II (\ie $d_{s}^{II}=d_{16}$=2.9\AA), depending on the shortest bond ($d_s$) of an irregular triangle which is formed by three Se atoms. $\text{TaSe}_3$ has two type I and two type II chains in a unit cell. The two chains of each type are related by
inversion symmetry. The four prismatic chains have strong bonds (\ie Se5-Ta1-Se3 and Se5-Ta2-Se6) along the $a+c$ direction, forming the TaSe$_3$ layers as depicted by green dashed lines. The interlayer hoppings are relatively weak compared with the intralayer hoppings, suggesting 2D $\text{TaSe}_3$
can be easily produced by exfoliation methods.

\begin{figure}[tp]
\includegraphics[width=3.5in]{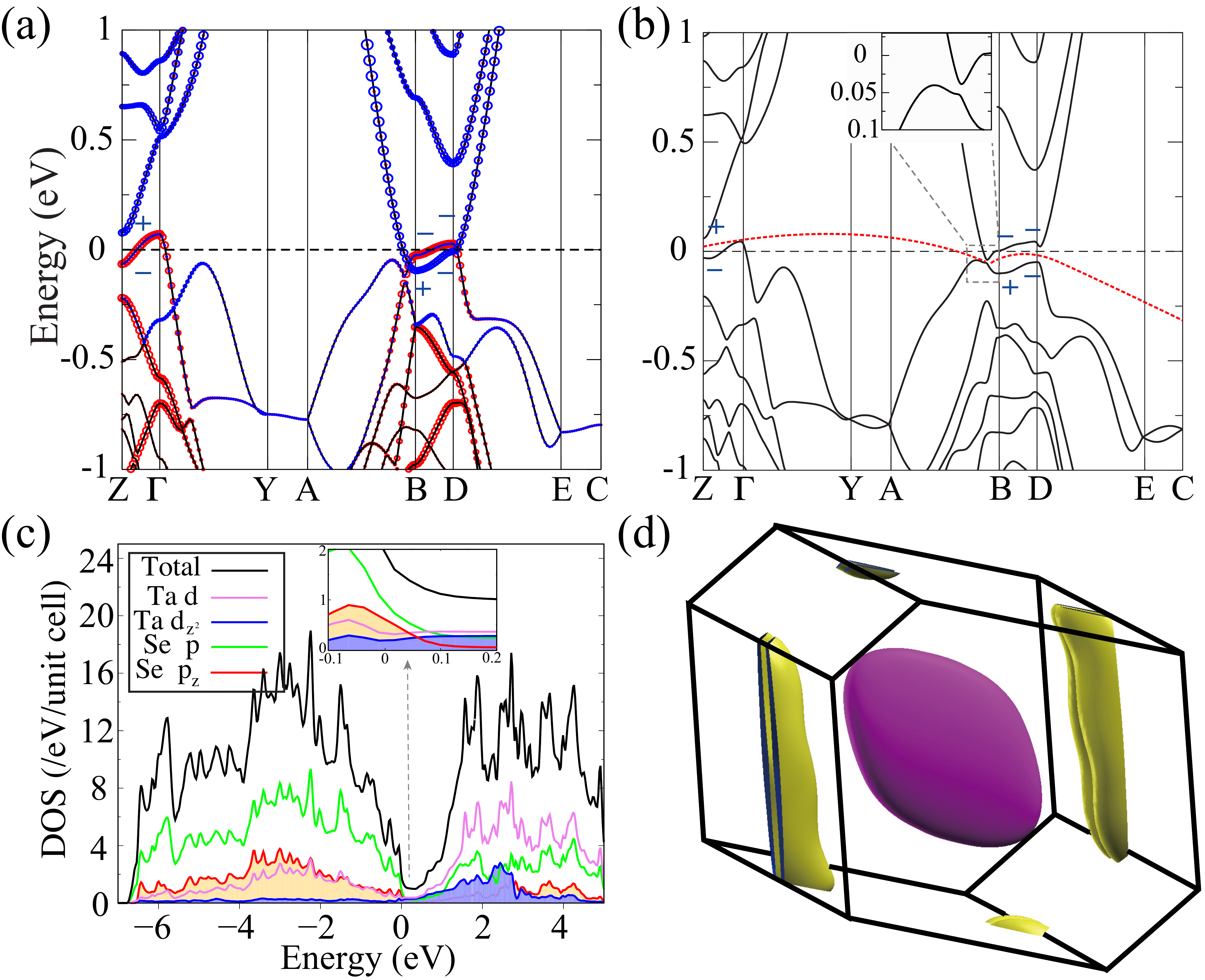}
      \caption{(Color online) Electronic structures of TaSe$_{3}$ without (a) and with (b) SOC. The calculated parity eigenvalues of highest valence band (HVB) and the lowest conduct band (LCB) at B and D points are given explicitly. The size of red and blue circles in (a) represents the weights of Se3 $p_z$ and Ta1 $d_{z^2}$, respectively. The inset in (b) shows the zoom-in band structure around the band crossing point located on the A-B line. The red dashed line in (b) indicates the existence of the continuous direct gap. (c) The calculated total DOS and PDOS of TaSe$_{3}$ . The inset in (c) shows the DOS around the E$_F$. (d) The Fermi surface of TaSe$_{3}$.}
\label{band}
\end{figure}

Next, we discuss the qualitative electronic structure of $\text{TaSe}_3$ by way of a simple Zintl-Klemm concept\cite{nesper2014zintl}. In
the type I chain of TaSe$_3$, where the distance between two Se atoms is short enough (\ie $d_s^I$=2.575 \AA), a strong covalent $p$-$p$
bond is formed. Thus, the oxidation state of the Se$_3$ triangle is ($\text{Se}^{2-}+\text{Se}^{2-}_2$). That's why ZrSe$_3$ is a semiconductor with Zr$^{4+}$ and only type I chains\cite{island2017electronics}. However, in the type II chain, the
distance is not short enough; namely, the $p$-$p$ bond is broken. These Se atoms exhibit the normal oxidation states (3Se$^{2-}$).
At the ionic limit, the chemical valence of tantalum is about 5+ ($\text{Ta}^{5+}$). Thus, (TaSe$_{3}$)$_4$ can be formulated as
2(4$\text{Se}^{2-}$+$\text{Se}^{2-}_2$+2$\text{Ta}^{5+}$) in a primitive unit cell. According to this electron counting model, TaSe$_3$
would be a semiconductor. However, $\text{TaSe}_3$ is metallic according to the transport measurements\cite{buhrman1980superconducting,bjerkelund1966011},
which implies band inversion may occur in the electronic band structure. In order to fully understand the semimetallic
propertes of TaSe$_3$, the first-principles calculations have been performed systematically.

\begin{figure}[tp]
\includegraphics[width=3.4in]{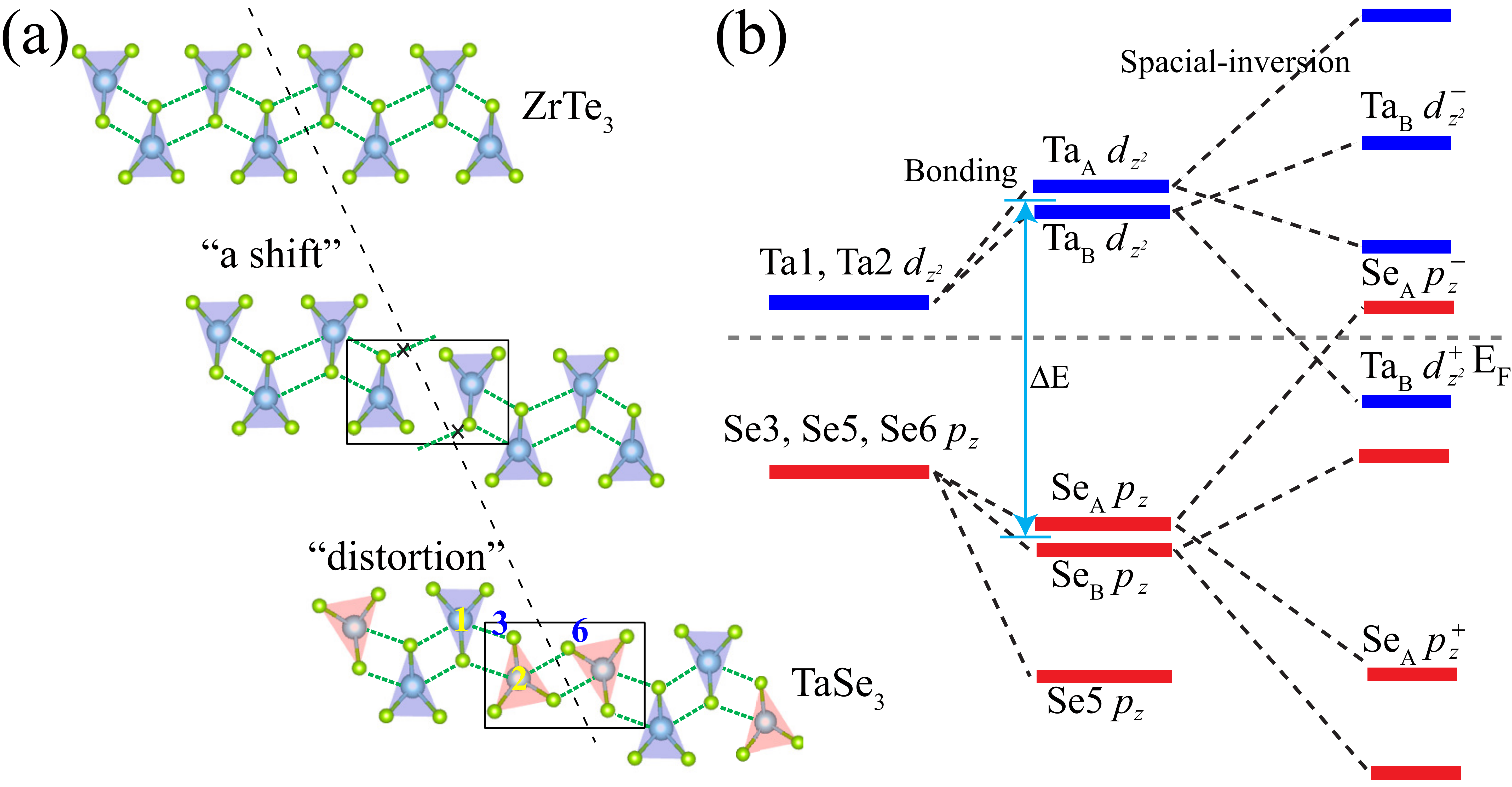}
\caption{(Color online) The evolution of the crystal structures and the schemetic diagram of the band-inversion mechanism. (a) The upper pattern is a layer of the ZrSe$_3$ structure, which is a semiconductor and only contains type I chain with Zr$^{4+}$, Se$^{2-}$ and (Se)$_2^{2-}$ states. The middle pattern is the intermediate structure generated by a shift in the $a$-direction (the black dashed line), which breaks the previous bonds of Se3 atoms, denoted by the symbol ``x". The lower pattern is a layer of the TaSe$_3$ structure. The distortion happens in the chains in the box, changing the type I chains to the type II chains. The new bonds between Se6 and Ta2 are built. (b) Schematic diagram of the band evolution in  TaSe$_3$, starting from the atomic orbitals $d_{z^2}$ of Ta1 and Ta2, and $p_z$ of Se3, Se5 and Se6.
}\label{inv}
\end{figure}

When SOC is ignored, the calculated band structure along high symmetric lines in the BZ is shown in Fig. \ref{band}(a). 
There is always a direct gap between the conduction bands and valence bands, except two crossing points on the AB and DE lines.
To elucidate the mechanism of the band inversion, we have calculated the projected weights
on six nonequivalent Se atoms (\ie $\text{Se1}, \text{Se2},\cdots$, and $\text{Se6}$, as shown in Fig.~\ref{structure}(b)) and two nonequivalent
Ta atoms (\ie $\text{Ta}1$ and $\text{Ta}2$, as shown in Fig.~\ref{structure}(b)), respectively. In the fatted-band plot of Fig.~\ref{band}(a),
we denote the weights of $\text{Se3}$ $p_z$ orbital ($\hat z || \vec b$) and $\text{Ta1}$ $d_{z^2}$ orbital by the size of the red and blue
circles, respectively. We can clearly see that the $p_{z}$ band and the $d_{z^2}$ band have an overlap at the Fermi level ($\text{E}_\text{F}$).
Further calculated results indicate that the up-going $d_{z^2}$ band mainly comes from the $d_{z^2}$ states of both $\text{Ta1}$ and $\text{Ta2}$,
consistent with the ligand crystal splitting of five $d$ orbitals in a prismatic cage. However, the down-going $p_z$ band is mostly from
the $p_z$ orbitals of $\text{Se3}$ and $\text{Se6}$ atoms (both of them belong to the ``broken" type II chains). To some extent, it suggests that the metallic band structure has to do with the type II chains. These similar results are also obtained by our calculated partial density of states (PDOS) in
Fig.~\ref{band}(c), which show that the Se $p$ states and Ta $d$ states are mainly located below and above the $\text{E}_\text{F}$, respectively,
with the hybridization between them. Near $\text{E}_\text{F}$,  the states are dominated by $p_z$ states of $\text{Se3}$ and $\text{Se6}$ atoms
and $d_{z^2}$ states of $\text{Ta1}$ and $\text{Ta2}$ atoms.

\begin{figure}[tp]
\includegraphics[width=3.4in]{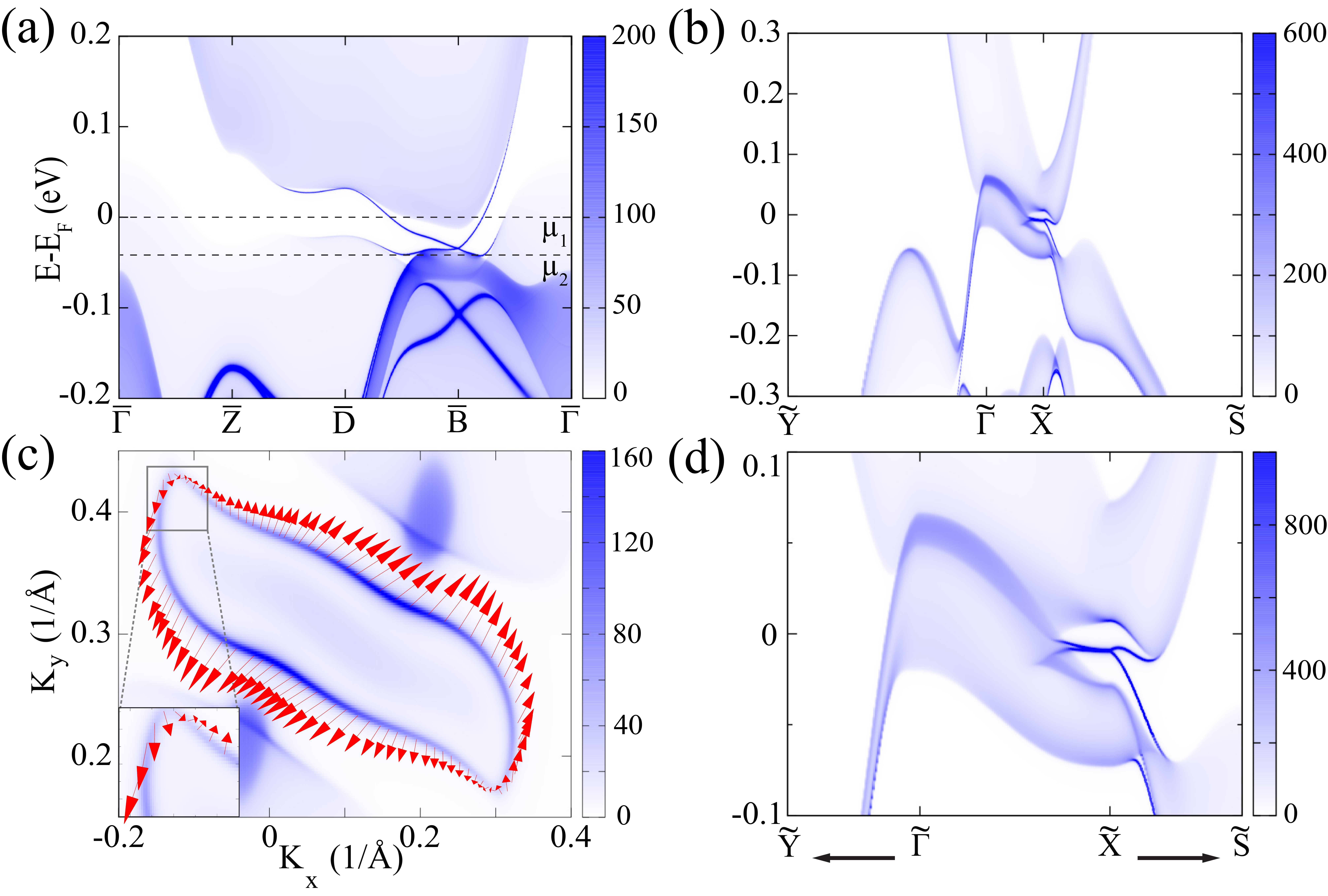}
      \caption{(Color online) Surface states of TaSe$_3$. (a) and (b) are the surface band structures of TaSe$_3$ on the (010) and (10$\bar1$) surfaces, respectively.
      The chemical potentials at Fermi level (E$_F$) and 40 meV below the Fermi level are represented by $\mu_1$ and $\mu_2$, respectively.
            (c) The constant energy contour (E=E$_F$) of the topological surface states and corresponding spin texture on the (010) surface of TaSe$_3$. The inset shows the zoom-in Fermi surface around the upper left corner. (d) Zoom-in surface band structures around the Dirac point on the (10$\bar1$) plane.}
\label{surf}
\end{figure}

To fully understand the metallic electronic band structure, we first investigate the crystal structure evolution beginning with the layered structure of ZrSe$_3$. In ZrSe$_3$ type I chains are arranged, as shown in Fig.~\ref{inv}(a), as a perfectly layered structure. It can be changed to the crystal structure of TaSe$_3$ by the following two steps. First, a shift may occur along the $a$-direction for every four prismatic chains (MX$_3$)$_4$, as shown in Fig.~\ref{inv}(a). As such, the bonds between selenium and tantalum atoms crossing the black dashed line are broken, as depicted by the symbol ``x" in Fig.~\ref{inv}(a). Second, the distortion can happen in the prisms of the box by breaking the shortest $p$-$p$ bonds between Se atoms. In addition, the new bonds between Ta2 and Se6 are built. Together, one can find that the crystal environment of Se3 and Se6 atoms of type II chains can change dramatically.

The band inversion can be understood from the bonds (green dashed lines) of Se5-Ta1-Se3 and Se5-Ta2-Se6, which support the layered structure of TaSe$_3$ (see Fig.~\ref{structure}(b) and Fig.~\ref{inv}(a)). In the atomic limit, the energy levels of the Ta $d$ orbitals are higher than Se $p$ orbitals. Under the crystal field of the prismatic cage, the $d_{z^2}$ orbitals are lower than other $d$ orbitals. The $p_z$ orbitals of Se atom are higher than other $p$ orbitals, since it doesn't orient toward the Ta atoms. Therefore, only $p_z$ orbitals of Se3, Se5 and Se6, and $d_{z^2}$ orbitals of Ta1 and Ta2 are considered in the schematic diagram of the band inversion, as shown in Fig. \ref{inv}(b). Starting from the atomic limit, the energy level of the Ta $d_{z^2}$ orbitals is higher than that of the Se $p_z$ orbitals. In step I, because Se5 forms two bonds while Se3 (or Se6) forms only one bond, the $p_z$ state of Se5 is pushed much lower than the $p_z$ state of Se$3$ (or Se$6$). The hybridization makes the two $p_z$ orbitals of Se3 and Se6 form two mixed states, called Se$_A$ and Se$_B$, respectively. Similarly, the two mixed states of Ta $d_{z^2}$ orbitals are called Ta$_A$ and Ta$_B$. The energy levels are illustrated in the middle of Fig.~\ref{inv}(b). In step II, the inversion symmetry is taken into consideration. Each state can split into two hybridized states, one bonding state and one antibonding state, according to the parity. The band inversion in TaSe$_3$ happens between the bonding state of Ta$_B$ and antibonding state of Se$_A$, which is consistent with our fatted band calculations and PDOS. This band inversion mechanism is further confirmed by the phase diagram under strains, as will be shown later.

\begin{figure}[tp]
\includegraphics[width=3.2in]{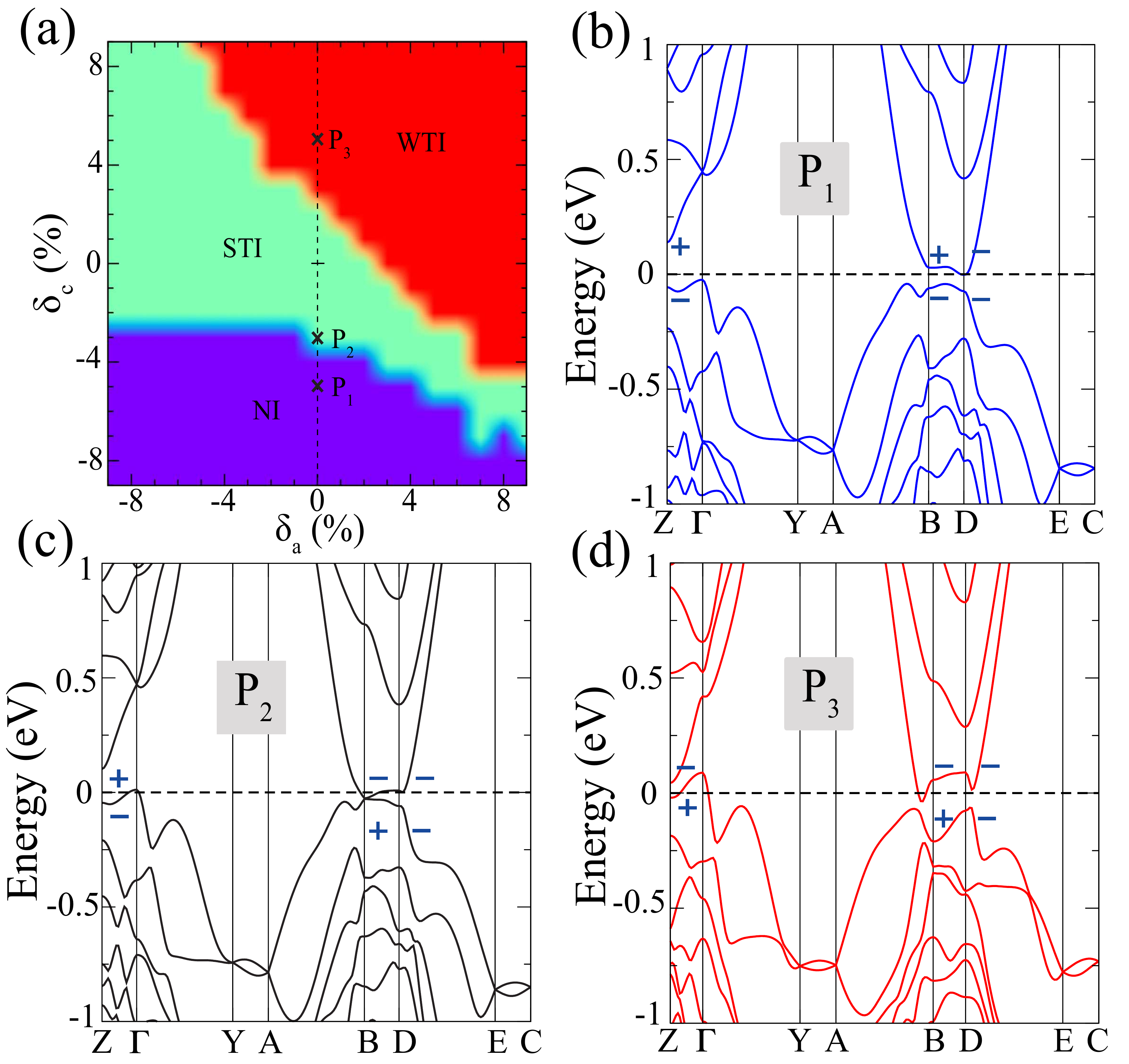}
      \caption{(Color online) Phase diagram of TaSe$_3$ with strain along both a- and c-directions. (a) The blue, green and red regions represent NI, STI and WTI phases, respectively.
(b), (c) and (d) are band structures of $\text{TaSe}_3$ with SOC for $c=0.95c_0$ (P$_1$), $c=0.97c_0$ (P$_2$) and $c=1.05c_0$ (P$_3$), respectively. The parities of HVB and LCB at three TRIMs (B, D and Z) are given.}
\label{evolution}
\end{figure}

The inclusion of SOC leads to gap opening at the band crossing points, as seen in the band structure of Fig.~\ref{band}(b). However,
the maximum of the valence bands is still higher than the minimum of the conduction bands, which gives rise to the semimetallic
properties of TaSe$_3$. The calculated Fermi surfaces consist of a large hole pocket enclosing $\Gamma$ point and two electron pockets
near B point, as shown in Fig.~\ref{band}(d). Further parity analysis indicates that the inverted two bands have different parity
eigenvalues at B point, while they have the same eigenvalue (\ie $-1$) at D point. Note that the band inversion has already happened
even without SOC. The band inversion is attributed to the unique structure of the type II chains, especially the Se3-Ta1 and Se6-Ta2 bonds, which is
distinct from the SOC-induced band inversion of the well-known Bi$_2$Se$_3$ family.

To classify TIs, one needs to compute four $Z_2$ topological indices $(\nu_0;\nu_1,\nu_2,\nu_3)$, where $\nu_0$ is
a strong topological index and $(\nu_1,\nu_2,\nu_3)$ are three weak topological indices. Since there is a continuous
direct SOC gap, the Fu-Kane $Z_2$ invariants are well defined for the occupied bands below the gap in TaSe$_3$.
Due to the existence of inversion center in $\text{TaSe}_3$, they can be easily calculated by Fu-Kane parity
criterion\cite{fu2007topological} at eight TRIM points ($\Gamma_i,i=1,2,\dots 8$).
The strong topological index is given by $(-1)^{\nu_0}=\prod_{i=1}^8\delta(\Gamma_i)$, where $\delta(\Gamma_i)$ is the product of parity
eigenvalues of the bands at $\Gamma_i$ without counting their time reversal partners. Three weak topological indices are defined at the
four TRIM points in a plane offset from $\Gamma$ point. Explicitly, $(-1)^{\nu_1}= \delta(A) \delta(B)\delta(D)\delta(E)$, $(-1)^{\nu_2}= \delta(A)
\delta(C)\delta(E)\delta(Y)$ and $(-1)^{\nu_3}= \delta(C) \delta(D)\delta(E) \delta(Z)$. The space group of $\text{TaSe}_3$ is non-symmorphic
with a screw symmetry $\bar C_{2b}$, which is a twofold rotation about the $\vec b$-axis followed by a half lattice translation in the same
direction ($\vec b/2$). It satisfies the relation $\bar C_{2b} I= t(\vec b)I \bar C_{2y} $, where $I$ is inversion symmetry and the lattice translation
$t(\vec b)$ can be represented by a phase factor in the Bloch basis: $t(\vec b)=exp(-i\vec b\cdot \vec k)$. Thus, at the TRIM points in the
$\vec k_2=\pi/b$ plane, \ie A, C, E and Y, with the expression $t(\vec b)=-1$, $\bar C_{2b}$ anticommutates with $I$. In combination of time
reversal symmetry, the anticommutating relation leads that all the states are four-fold degenerate, consisting of two parity $+1$ bands and two
parity $-1$ bands (\ie ``$++--$"). Considering that the total number of valence bands is $68=4\times 17$, we get $\delta(A)$=$\delta(C)$=$\delta(E)$=$\delta(Y)$=$(-1)^{17}$=$-1$.
Therefore, the strong topological index $\nu_0$ is determined by band inversion at other four TRIMs (Z, $\Gamma$, B and D). Their parity
products are calculated to be $\delta(Z)$=$\delta(\Gamma)$=$\delta(D)$=$-1$ and $\delta(B)$=$1$, which are extracted from standard numerical calculations based on first-principles calculations. Therefore, the $Z_2$ topological
invariants of $\text{TaSe}_3$ turn out to be (1;100).

In view of the fact that the hallmark of topological non-trivial property is the existence of topological non-trivial surface states,
the tight-binding Hamiltonians of semi-infinite samples are constructed by the maximally localized Wannier functions (MLWFs) for
all the Ta $d$ and Se $p$ orbitals, which are generated from the first-principles calculations. The surface Green's functions of the
semi-infinite sample are obtained using an iterative method\cite{marzari2012maximally}. The local DOS (LDOS), extracted
from the imaginary part of the surface Green's function, is used to analyze the surface band structures.
For STIs, the existence of an odd number of Dirac cones on the surface is ensured by the strong topological index $\nu_0$=1.
On the (010) surface of TaSe$_3$, a Dirac cone is obtained at $\bar{B}$ point in Fig. \ref{surf}(a), which hosts helical spin texture
at the energy contour (E=E$_F$) in Fig.~\ref{surf}(c). On the (10$\bar{1}$) surface, a Dirac cone is found at $\tilde{X}$ point, as shown
in Fig.~\ref{surf}(b) and its zoom-in plot in Fig. \ref{surf}(d). The existence of a single Dirac cone on the surfaces is consistent
with the STI phase.

\begin{figure}[tp]
\includegraphics[width=3.2in]{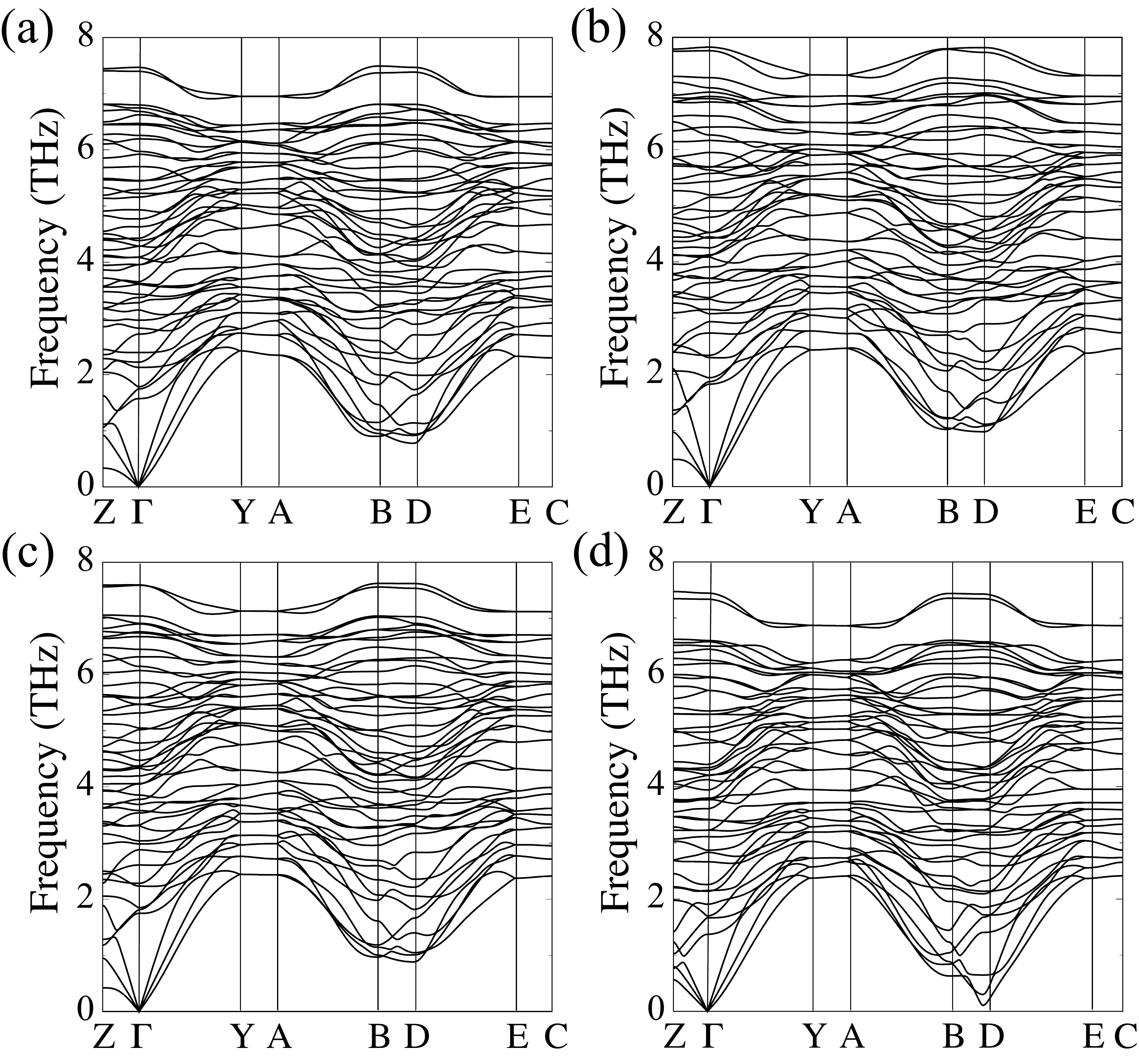}
      \caption{(Color online) Phonon dispersion of TaSe$_3$ with c=1.0 c$_0$, c=0.95c$_0$ (P1), c=0.97c$_0$ (P2) and c=1.05c$_0$ (P3), as shown in (a), (b), (c) and (d), respectively.}
\label{phonon}
\end{figure}

Since it's a layered structure with trigonal-prismatic chains going in the $b$-direction, the lattice parameters in $a$- and $c$-directions
are supposed to be more sensitive to temperature or strain than that in the $b$-direction. Therefore, the strain effects along the
$a$- and $c$-directions have been systematically investigated in our calculations, in order to check the stability of the STI phase.
For each given lattice constant, the internal positions of the atoms are fully relaxed until the force on each atom satisfies the
required precision. The computed phase diagram is shown in Fig. \ref{evolution}(a) as a function of lattice parameters
$a$ (horizontal axis) and $c$ (vertical axis). On the vertical dashed line in Fig. \ref{evolution}(a) (\ie the strain in the
$a$-direction is zero), when $c$ grows from $0.95c_0$ to $1.05c_0$ (with $c_0$ the lattice parameter without strain), it changes from
a normal insulating (NI) phase, to STI phase, and then to WTI phase due to band inversion happening at TRIMs successively.
Explicitly, at the beginning ($c=0.95c_0$), no band inversion is found in Fig. \ref{evolution}(b) and the system is a trivial
insulator. When $c=0.97c_0$, two bands of opposite parity become touching at B point, as shown in Fig. \ref{evolution}(c), resulting
in a Dirac semimetal phase at the phase transition point. After the transition point, the system enters STI phase.
By increasing $c$ further ({\it e.g.} $c=1.05c_0$), another band inversion occurs at Z point, the system turns to WTI
phase, as shown in Fig. \ref{evolution}(d).
As we can see in Fig. \ref{evolution}(a), both STI (green) and WTI (red) phases
survive in a considerable region compared with the NI (blue) phase. The phase boundaries are Dirac
semimetals, which may also be observed in experiments. The phase diagram provides us an important roadmap to regulate
the topological phase transitions in $\text{TaSe}_3$.
Compared with the reported topological materials, TaSe$_3$ has two unique features. First, the compressive strain can remove the band inversion, which is different from the common concept that the compressive strain usually enhances the band inversion\cite{nie2015strain,sisakht2016strain,shao2017strain}. 
The compressive strain shortens the distances of the bonds of Ta1-Se3 and Ta2-Se6, the hybridization becomes stronger and the average gap $\Delta E$ is enlarged between Se$_A$ $p_z$ (Se$_B$ $p_z$) and Ta$_A$ $d_{z^2}$ (Ta$_B$ $d_{z^2}$) states. 
Second, {\color{blue}many TIs} have an $sp$-type band inversion or a $pp$-type band inversion \cite{nie2015strain,sisakht2016strain,shao2017strain}, while TaSe$_3$ has a $pd$-type band inversion. Third, most TIs only have band inversion at one TRIM point (such as the $\Gamma$ point)\cite{zhang2010first,nie2015quantum}, while TaSe$_3$ has band inversion at three TRIM points,  {\color{blue}which leads} to fruitful nontrivial phases under different strains, including 3D STI, 3D WTI and Dirac semimetal.

Next, we would like to discuss the possible CDW in TaSe$_3$. It is well-known that no CDW is observed in NbS$_2$. However, the
phonon dispersion shows imaginary frequencies based on first-principles calculations \cite{heil2017origin}.
In order to exclude the possibility of the $``$latent$"$ CDW instability and greatly support our predictions of nontrivial topological properties in TaSe$_3$, we carefully calculate the phonon dispersion for TaSe$_3$, which is shown in Fig. \ref{phonon}(a). No imaginary frequency is found in the dispersion. The phonon calculations have also been checked with respect to many parameters, such as $k$-point grid, type of smearing and size of supercell etc. In addition, we also calculate the phonon dispersion of TaSe$_3$ with c=0.95c$_0$ (P1), c=0.97c$_0$ (P2) and c=1.05c$_0$ (P3), as shown in Fig. 6(b), 6(c) and 6(d), respectively. No imaginary frequency can be found in any of them. Therefore, it is safe to exclude the effect of CDW in TaSe$_3$.

\section{DISCUSSION and Conclusion}
To achieve topological superconductivity, it is challenging to find a material with both nontrivial topology of the electronic structure and superconductivity.
A prior theoretical proposal of intrinsic non-trivial material is high-T$_c$ superconductor FeTe$_x$Se$_{1-x}$~\cite{wang2015topological} with $x=0.5$ supporting topological surface states at $\text{E}_\text{F}$. Very recently, the topological superconducting phase has been verified on (001) surface by spin-resolved ARPES experiments~\cite{zhang2017observation}. Here we show another promising candidate TaSe$_3$, whose superconductivity has been reported for many years\cite{sambongi1977superconductivity,yamamoto1978sc,nagata1989superconductivity}, is topologically non-trivial and has spin-momentum locking states on the surface. Compared to the FeTe$_x$Se$_{1-x}$ system, TaSe$_3$ does not require doping to introduce superconductivity, implying this superconductor can be grown in high-quality single crystals.
At the Fermi level ($\mu_1$), as we show in Fig.~\ref{surf}(c), the surface states are well separated from the bulk states. Based on Fu-Kane's proposal, the superconducting states of the surface states induced by the bulk superconductivity below T$_c$=2.3 K can be topologically nontrivial. However, if the chemical potential lies 40 meV ($\mu_2$) below the Fermi level in Fig.~\ref{surf}(a), the surface states merge into the bulk states, which could kill the topological superconducting state and make the surface topological superconducting state sensitive to the position of the chemical potential. The future experimental work is needed to search for the potential topological superconductivity in the system.

In conclusion, we have calculated the electronic structure, and topological properties of $\text{TaSe}_3$ with $P2_1/m$ crystal
structure by using density functional theory. The calculated topological invariants are (1;100), which indicate that it belongs
to the STI phase. A single Dirac cone is obtained in the calculated surface spectra. Further systematical calculations of strain
effects suggest $\text{TaSe}_3$ can realize multiple topological non-trivial phases under strains, including STI, WTI and
Dirac semimetal phases. These topological non-trivial phases survive in a wide range of the strain along the
$a$- and $c$-axes, making $\text{TaSe}_3$ an attractive  platform to study the topological phase transitions.
In addition, the co-existence of the topological phases and superconductivity of $\text{TaSe}_3$ suggests that it could be a
realistic system to study the interplay between topological and superconducting phases in the future.

\begin{acknowledgments}
We thank Prof. Robert J. Cava for useful discussions. F. B. P. and S. N. were supported by Stanford Energy 3.0.
W. X. was supported by LSU-startup funding and the Louisiana Board of Regents Research Competitiveness
Subprogram (RCS) under Contract Number LEQSF (2017-20)-RD-A-08. Z. W. was supported by the Department
of Energy Grant No. DE-SC0016239, the National Science Foundation EAGER Grant No. NOA-AWD1004957,
Simons Investigator Grants No. ONR-N00014-14-1-0330, No. ARO MURI W911NF-12-1-0461, and No. NSF-MRSEC DMR- 1420541,
the Packard Foundation, the Schmidt Fund for Innovative Research, and the National Natural Science Foundation of China (No. 11504117).
L.X. and R.J. are supported by Department of Energy Grant No. SC00i6315.
\end{acknowledgments}

\bibliography{TaSe}

\end{document}